# Time-Transient Wireless RF Sensor with Differentiative Detecting Capability for Target Ionic Solution of Water and Dielectric Objects Introduced into Water

Sobhan Gholami, Emre Unal, and Hilmi Volkan Demir, *Fellow Member, IEEE*

**Abstract**— A novel wireless microstrip-based RF sensor designed for detecting changes in ionic content of water and the addition of solid contaminant objects is proposed and demonstrated. The sensor can be installed on the exterior wall of dielectric containers and customized according to the material of the container to enable wireless sensing. Its operation within the lower microwave frequency range (670-730 MHz) serves to minimize signal attenuation in water and streamlines circuitry design. The most significant feature of this sensor is its unique design, rendering it impervious to its surrounding environment. This not only shields it from environmental noise but also maximizes its sensitivity by efficiently utilizing incoming power for sensing purposes. The sensor exhibits remarkable sensitivity, capable of detecting solute concentrations as low as $3.125 \times 10^{-3}$ M in water. It can also detect the insertion of foreign solid objects into the container from the exterior wirelessly and distinguish them from liquids being added. As a proof-of-concept demonstration, the sensor in this study was optimized for a porcelain wall of 10-12 mm thickness. The sensor's small size and the materials used for its fabrication make it adaptable to a wide range of applications in industries such as food, pharmaceuticals, and bathroom fixtures. The aforementioned properties position the sensor as an ideal choice for various smart bathroom applications, where accurate and reliable water use monitoring is essential for efficient water conservation.

**Index Terms**—Microstrip patch antenna, microwave propagation, microwave sensors, chemical and biological sensor, signal analysis, water conservation, water monitoring, green cleaning.

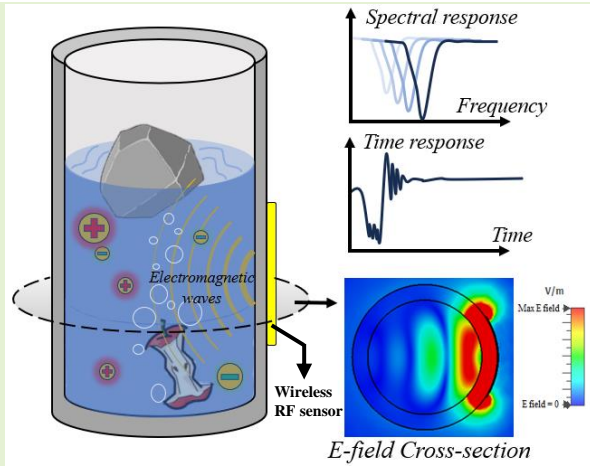

## I. Introduction

In recent years, the pressing issue of global warming and its effect on water resources [1], [2], [3],has attracted much attention in using technology for water conservation. There have been efforts in the reduction of water use, especially in bathrooms, as a significant source of water consumption at homes by imposing regulations [4]. In addition, new toilet designs have been introduced that require less water for flushing [5], [6]. Sensors have also been employed to promote water conserving practices in bathroom facilities. Examples of such sensors are IR proximity sensors for automatic urinal flushing in public bathrooms, as well as load detectors installed under toilet lids for automatic flushing [7], [8].

While IR sensors activate the flushing mechanism based on the reflection of electromagnetic wave (EMW) of a specific wavelength from the presence of an object or person in front of the sensor, in load sensors, mechanical energy is converted into electrical signals to show the presence of a person sitting on the lid and obviate the need to mechanically press the flush button or press it multiple times unnecessarily after standing up. Yet non operates based on the actual presence of excreta in the toilet.

To address the sanitation concerns, it is advisable to utilize electromagnetic waves for sensing in bathrooms. EMW's do not require physical contact with human excreta and thus minimize unwanted contamination. A considerable number of

This paper was submitted for review on 25.01.2024 This project is supported by TUBITAK's Industry Innovation Network Mechanism (SAYEM) Program under grant no. 121D010 (Akıllı Ev Platformu).

S. Gholami is a graduate student with the department of Electrical and Electronics Engineering, Bilkent University, TR-06800, Ankara, Turkey (e-mail: gholami@ee.bilkent.edu.tr)

E. Unal is with the Institute of Materials Science and Nanotechnology, Bilkent University, TR-06800 Ankara, Turkey (e-mail: unale@bilkent.edu.tr )

H. V. Demir is with the Department of Electrical and Electronics Engineering and the Department of Physics, Institute of Materials Science and Nanotechnology (UNAM), Bilkent University, TR-06800 Ankara, Turkey, and also with the School of Electrical and Electronic Engineering and the School of Physical and Mathematical Sciences, Nanyang Technological University, Singapore 639798 (e-mail: volkan@bilkent.edu.tr ).

research studies have been conducted in microwave frequencies and demonstrated the potential in microwave devices being employed as sensors [9], [10], [11] [12], [13], [14]. These sensors have demonstrated to detect the alterations in the electromagnetic properties of a wide range of materials. They employ reflection, reflection/transmission, and resonance perturbation measurement methods to achieve sensing in the framework of 1 and/or 2 port networks. Nevertheless, a significant proportion of these sensors rely on direct contact between the material under test (MUT) and the sensor. This was observed using open-ended coax cables [15], [16], [17], or microstrip resonators [18], [19], [20]. Research has also been conducted into confining the MUT in microfluid tubes to isolate the MUT; however, the restricted size of the tubes and delicate operation of the sensors limit their applications in a bathroom environment [21], [22], [23].

Antennas, in a number of cases, were utilized to discern the variations in the electromagnetic properties of materials from the transmission information [24], [25], [26]. Still, the application of antennas as a sensing mechanism is only effective when the MUT is completely separated, which does not seem feasible in a restroom environment where various variable elements exist. The reflection response of an antenna has been used for determining concentrations of sugar and salt in water in [27] where the antenna was operated as a resonator and immersed in the solution under test. In [28], [29], an ultra-high-quality factor resonator was shown. The mentioned resonator is capable of distinguishing between different liquids inside a tube submerged in water in the near field, primarily due to its exceptionally large quality factor. However, this was only presented for the cases where the layer between the resonator and the MUT was extremely thin to allow the field to penetrate inside. A two-port microstrip resonator was introduced in [30] capable of detecting different layers of materials (mainly for oil and sand) wirelessly based on variations in the transmission response of the resonator, employed to make out the difference in the state of materials under test. One drawback in such sensors is that the electromagnetic wave does not propagate inside the medium, making them unsuitable for detecting objects located farther inside the medium.

This paper presents a novel time-dependent single-frequency sensor design capable of detecting changes in the concentration and presence of dielectric objects within water inside a porcelain container. This capability has the potential to play a distinct role in determining the necessary amount of water for each flushing, thereby preventing the waste of water from excessive flushing in toilets. The operational frequency of the sensor is carefully selected to minimize the loss due to attenuation of electromagnetic waves while also ensuring that the wavelength is comparable to the size of the objects inserted into the water to carry out detection.

The unique characteristics of this design allow it to be installed on the outer surface of a toilet and perform sensing without being exposed to human excreta. Moreover, the distinctive design of this resonator renders it completely uncoupled from its surroundings, establishing a sensing environment where the only variables are confined within the toilet. Additionally, this design is compact in size, making it compatible with existing toilet setups.

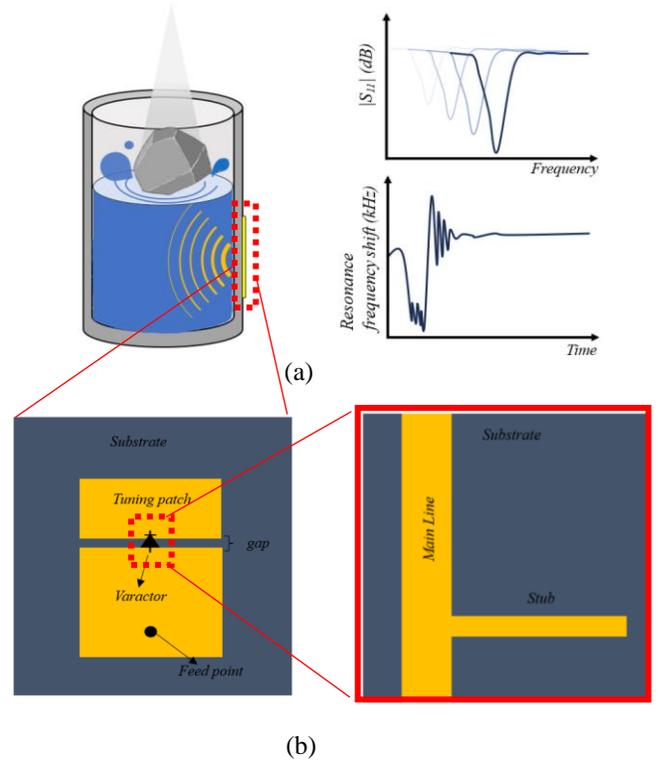

Fig. 1. Conceptual illustration of wireless RF sensing in time-transient mode: (a) Conceptual sketch, (b) Replacing the varactor in a frequency reconfigurable directive patch antenna with a transmission line and a stub to create a capacitive element.

## II. DESIGN OF THE PROPOSED SENSOR

The sensor's desired characteristics pose a significant challenge in the design process. Achieving the right balance between size, directivity, conformity, and operating frequency is a formidable obstacle. Microstrip patch configurations, with their inherent qualities of conformity, low profile, and versatility, emerge as the optimal choice for sensor design. The antenna, as depicted in Fig. 1, successfully meets the conformity and directivity criteria for our application. Nevertheless, a notable challenge arises due to its resonance frequency, which significantly exceeds the intended operational range (below 1 GHz) [31]. Furthermore, the integration of varactors in the design introduces additional DC circuitry demands, further complicating the final product— an aspect that is less desirable.

To lower the resonance frequency while maintaining directivity, we removed the varactor and strategically incorporate capacitive elements between the main patch and the tuning patch, as shown in Fig. 1. In this design, open-ended microstrip transmission lines, also referred to as stubs, are positioned along another transmission line connecting the main patch to the tuning patch. These stubs are tuned to serve as capacitive elements.

According to the two-port model of a terminated transmission line, looking into a transmission line toward load, the input impedance [32] is:

$$Z_{in} = Z_0 \frac{Z_L + jZ_0 tan\beta l}{Z_0 + jZ_L tan\beta l} \quad (1)$$

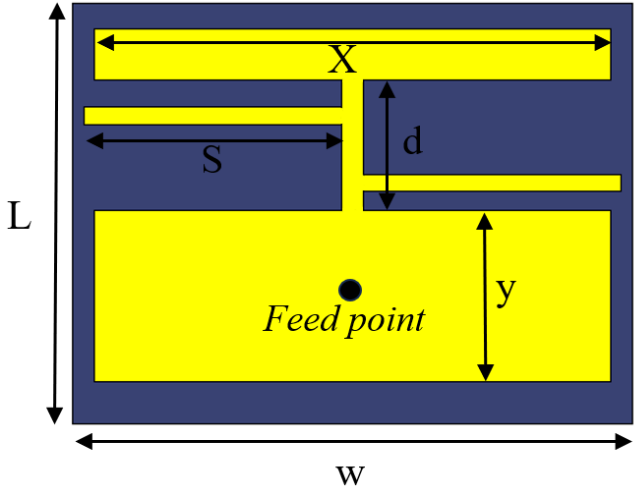

Fig. 2. Proposed sensor's structure. W=65 mm, L=50 mm, d= 15.2 mm S= 30 mm, x=60 mm, y=23 mm.

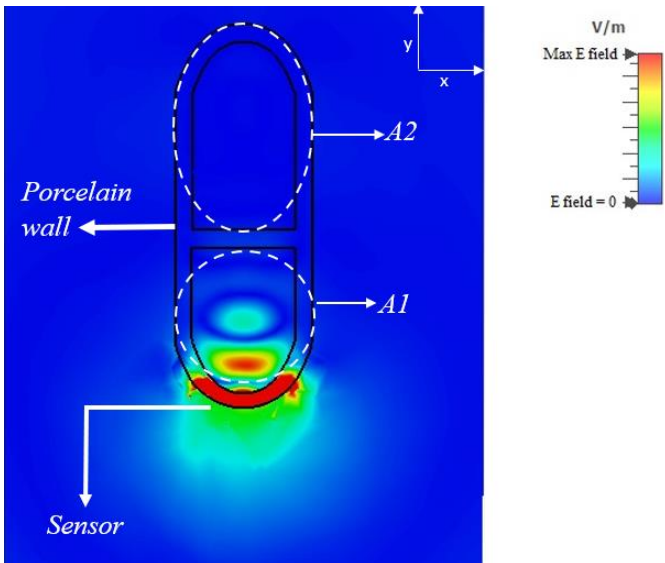

Fig. 3. Z-cut representation of propagation of EMW through porcelain into water. A1 is the front side of the S pipe underneath the toilet bowl, where the excreta are first inserted. A2 is in the outgoing part of the S pipe.

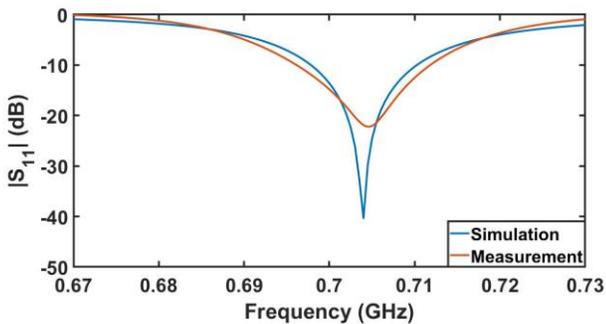

Fig. 4. Experimentally measured reflection coefficient compared to numerical solution of the fabricated sensor.

where $Z_0$ represents the intrinsic impedance of the line, $Z_L$ the terminating load and $l$, the length of the line.

Setting $Z_L = \infty$ for an open-ended transmission line makes:

$$Z_{in} = -jZ_0 cot\beta l \tag{2}$$

According to (2), the open-ended transmission line can provide almost any reactance by tuning the length and the intrinsic impedance of the line. The intrinsic impedance of the line is determined using [33] :

$$Z_0 = \frac{120\pi}{\sqrt{\epsilon_{eff}}\left[\frac{w}{h} + 1.393 + 0.667\ln\left(\frac{w}{h} + 1.44\right)\right]} \tag{3}$$

$$\epsilon_{eff} = \frac{\epsilon_r + 1}{2} + \frac{\epsilon_r - 1}{2}\left[1 + 12\frac{h}{w}\right]^{-1/2} \tag{4}$$

In (3) and (4), w is the width of the stub; h is the height of the underlying dielectric substrate and $\epsilon_r$ is the relative dielectric constant of the substrate.

To implement the stub, a 50-ohm microstrip transmission line was drawn between the patch and tuning patch. Two open-ended stubes were introduced on both sides of the line to provide capacitance. The location of the stubs from the tuning patch, their length and length of the transmission line was tuned using Computer Simulation Technology Microwave Studio to achieve the lowest resonant frequency possible. The sensor was designed on Rogers corporations RT5880 double sided copper with dielectric thickness of 0.79 and copper thickness of 0.018 mm shown in Fig. 2.

As the sensor's primary objective is to transmit EMW through porcelain into water, the tuning and optimization process was conducted within a porcelain medium immersed in water. This approach closely emulates the sensor's real operational conditions. By adopting this optimization method, reflections from the sensor/porcelain interface were minimized, allowing a higher proportion of the source power to penetrate the medium and facilitate sensing. Fig. 3 visually represents the electric field radiated by the sensor into the medium, demonstrating the sensor's ability to remain isolated from its surrounding environment—a key accomplishment sought in our design.

The design was physically realized on a double-sided copper laminate of Rogers corporations RT5880 (with $\epsilon_r$=2.2 and tanδ=0.009). Experimental testing was carried out on a porcelain container with a thickness between 10 to 12 mm. A comparison between numerical simulations and measurement results are presented in Fig. 4 where acceptable agreement between the numerical solutions and measurements is achieved.

## III. RESULTS AND DISCUSSION

The proposed sensor is capable of detecting the change in the concentration of water and sense the insertion of dielectric objects in the toilet. To evaluate the capability of the designed sensor for practical application, a special structure is designed to emulate the actual situation in a bathroom. The sensor is hot glued to the front section of the S pipe on the bottom of the porcelain toilet setup and it communicates with an Agilent technologies E5061B ENA through a SMA cable shown in the inset of Fig. 5. Careful attention was given to thoroughly remove all air gaps between the sensor and the porcelain to prevent multiple reflections between the two surfaces.

To test the sensor and determine its sensing range in response to changes in the ion content of water, we prepared samples of NaCl solutions with different concentrations in the range of $3.125 \times 10^{-3}$ to $5.000 \times 10^{-1}$ M (mol/L) using 99.9% pure NaCl. Additionally, we prepared an artificial urine solution following the formulation provided in [34] and maintained it at 37 °C. This allowed us to determine which concentrations of NaCl solutions at room temperature yield an equivalent electrical response when added to water instead of artificial urine.

In order to emulate the addition of feces to the medium, samples were prepared using the formulation in [35], which weigh around 45-50 gr each.

### A. Detecting the change in ionic concentration of water

The electrical response of a material to the incident electromagnetic wave is determined by its dielectric constant. The dielectric constant of pure water is well described by Debye model as a function of the frequency [36]:

$$\epsilon_w(f) = \epsilon_w(\infty) + \frac{\epsilon_w(0) - \epsilon_w(\infty)}{1 + i2\pi f \tau_w} \qquad (5)$$

where $\epsilon_w(\infty)$ is the dielectric constant of water at very high frequency, $\epsilon_w(0)$ is the dielectric constant of water at DC and $\tau_w$ is the relaxation time of pure water. Since (5) spans a broad frequency spectrum, it is reasonable to assume that the dielectric constant of water remains constant within the operational frequency range of this work (670-730MHz).

When ions are dissolved in pure water, the DC dielectric constant of water and relaxation time are influenced by the presence of ions, resulting in modifications to the relation described in (5) [37]:

$$\epsilon_s(f) = \epsilon_w(\infty) + \frac{\epsilon_s(0) - \epsilon_w(\infty)}{1 + i2\pi f \tau_s} - i\frac{\delta}{2\pi f \epsilon_0} \qquad (6)$$

Here $\epsilon_0$ is the dielectric permittivity of free space, $\epsilon_s$ is the static dielectric constant, $\tau_s$ is the relaxation time, and $\delta$ is the conductivity of the resultant solution, which are functions of the concentration of the dissolved ions. Detailed relationships governing these properties can be found in [37], [38], [39].The introduction of an ionic solution with a concentration higher than that of water alters the ion content of the water, leading to changes in $\epsilon_s$. These changes manifest as variations in the electrical response of the solution which in the case of this study, is the changes in the frequency of minimum reflection coefficient (S11).

To examine changes in the sensor's steady-state reflection coefficient, the toilet was filled with water containing various concentrations of NaCl. We analyzed the resulting resonance frequency shift $(f - f_0)$ concerning the baseline case when the medium consisted of only water $(f_0)$. These comparisons are visually represented in Fig 5. It is clearly seen that the increase in the ion content of the medium, leads to a consistent shift in the resonance frequency.

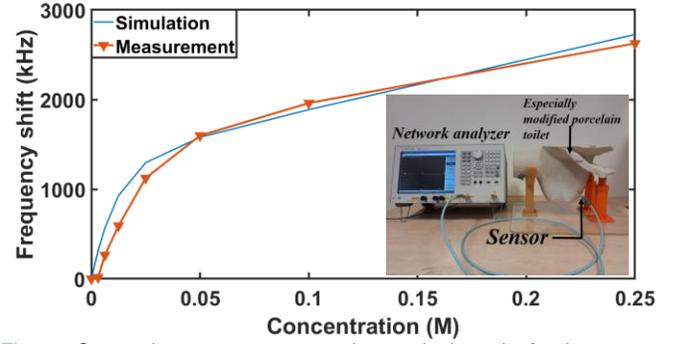

Fig. 5. Comparing measurement and numerical results for the frequency shift corresponding to various concentrations of NaCl in the measurement setup in the inset.

In the next step to observe the transient sensor response to the addition of ions, 200 ml volumes of solutions with concentrations of $3.125 \times 10^{-2}$ , $6.25 \times 10^{-2}$ , $1.25 \times 10^{-1}$, $2.5 \times 10^{-1}$ , $5 \times 10^{-1}$ M and artificial urine were slowly injected into the test toilet setup, simulating the introduction of urine into the medium. Fig 6. shows the shifts in the frequency *vs* volume of added solution for different solution concentrations. Horizontal dashed lines show the concentration of the medium. Step by step, as more solution was added to the medium while the volume was constant, the shift in the frequency monotonously increased.

Since we expect an exponential response from the addition of ions into the system, we fitted the data points using:

$$\Delta f = a(1 - e^{-bv}) \qquad (7)$$

Here a represents the maximum frequency shift expected to occur in steady state and b determines the rate at which the steady state is reached. v is the volume of solution in mL added.

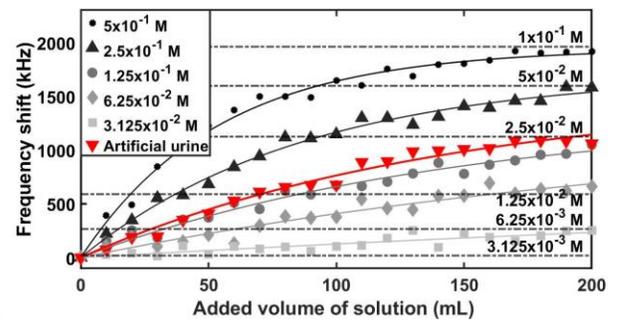

Fig. 6. Measurement results of changes in resonance frequency (frequency shift) with respect to the volume of added NaCl solution with different concentrations. Horizontal dashed lines show the frequency shift corresponding to concentration of the mixture inside the toilet.

## B. Simultaneously detecting the addition of solid objects into water in contrast to the ionic solution

The introduction of solid objects into a medium, when compared to the addition of solutions, often leads to significant and erratic shifts in resonance frequencies. Therefore, the transient response becomes a crucial aspect to consider. In the context of our study, when a solid object is introduced into the water, it generates ripples on the water's surface, with the amplitude of these ripples directly reflecting the energy associated with the inserted object. Consequently, analyzing the amplitude of these ripples serves as a valuable means of determining the presence of feces within the toilet. The sensor, loaded with water in design, captures and registers the ripples produced, which, in turn, induce direct changes in its resonance frequency. As a consequence, these ripples manifest as distinctive features on the frequency shift curve over time. Analyzing the frequency content of this time-dependent frequency shift curve offers distinctive features to differentiate feces from ionic solution.

To isolate the ripples from the frequency shift curve, we performed an instantaneous differential operation on the frequency shift data. This operation effectively eliminates any influence from the presence of solid objects within the sensor's electromagnetic field. Following this step, we employ Fourier transformation to obtain the frequency components within the derivative of the frequency shift curve.

For the Fourier analysis, it is important to consider the sampling time of the signal. In our study, we set a consistent sampling time of 110 ms between each data point. Fig. 7 illustrates the three critical stages of detecting, analyzing and comparing the ripples in the solid object insertion case together with the liquid injection case. For the most comprehensive examination, we conducted tests involving liquid injection from a height of 80 cm above the water level at a rate of 17 mL/s [40]. In Fig. 7(a), we present the frequency shift curve resulting from the introduction of three 50-gram pieces of artificial feces compared with injection of 220 mL of $1.25 \times 10^{-1}$ M of NaCl. Fig. 7(b) showcases the derivative of the frequency shift curve ($\frac{d(f-f_0)}{dt}$) for both these cases, and in Fig. 7(c), we present the Fourier transform of the frequency shift curves. Two prominent peaks within the range of 1.6 to 3.0 Hz are clearly observed in our analysis of solid insertion case. These peaks correspond to the dominant period of water ripples due to the depth of water in the toilet [41]. Significantly, these peaks consistently show up with noticeable magnitudes whenever a solid object is introduced into the water.

As it is expected (Fig.7(c)) ripples generated by the injection of liquids into the toilet exhibit substantially smaller magnitudes within the modulation frequency range of 1.6 to 3.0 Hz compared to instances involving solid object insertion. Naturally, feces displace greater amounts of water leading to higher amplitudes of waves formed on the surface. This allows to differentiate between the insertion of feces and urine into the toilet via analyzing the spectrum of the filtered data shown in Fig. 7(b). We achieve this by applying a bandpass filter and conducting spectral analysis on the frequency content of the derivative curve, ($\frac{d(f-f_0)}{dt}$), at regular time intervals. Following this approach also enables us to track the magnitude of the peaks within the $1.6 - 3.0$ Hz range over time.

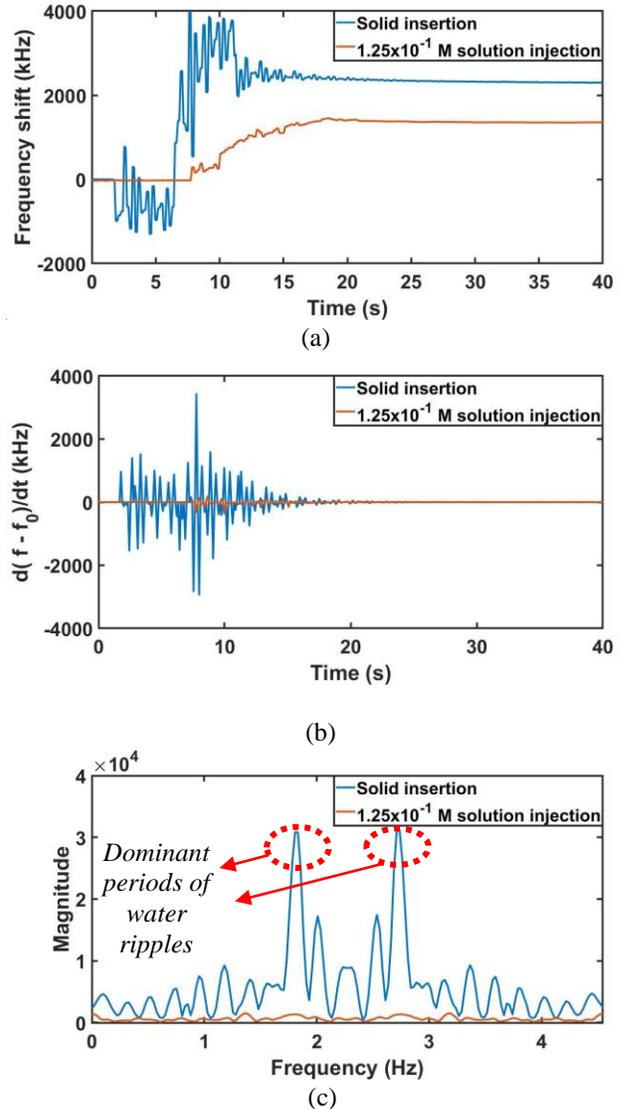

Fig.7. Measurement and analysis of injection of $1.25 \times 10^{-1}$ M solution with insertion of solids into the medium: (a) frequency shift curves corresponding to both cases, (b) derivative of the frequency shift curves in both cases, and (c) frequency content of the derivative of frequency shift curve.

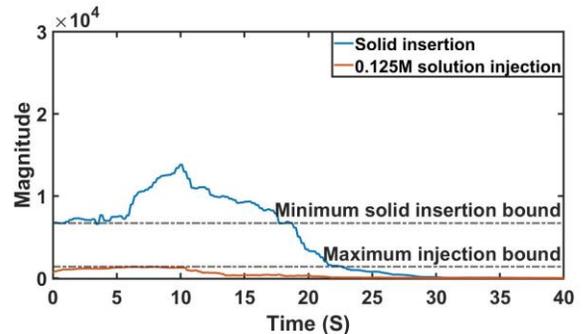

Fig. 8. Comparing minimum and maximum detected sensing signal magnitude of solid insertion and $1.25 \times 10^{-1}$ M NaCl solution injection.

In Fig. 8, we selected the scenario with the lowest magnitude out of 9 tests for solid insertion and compared it to the injection case where the highest magnitude was reached. This comparison reveals a significant difference between these two extreme scenarios, underscoring the reliability of this analysis method in distinguishing between the incoming of feces and urine in the toilet.

In this case study, the sensing algorithm is structured as follows: initially, the system conducts a comprehensive scan to identify the presence of solid insertions, promptly executing a washing procedure upon detection. The performed washing also removes any potential ionic solutions. In instances where no solid insertions are detected, the system shifts its focus to analyzing the resonance frequency's shift, specifically corresponding to the introduction of electrolytes. This analytical approach enables the system to discern the concentration of the added solution, empowering it with the capability to execute programmable actions commensurate with the concentration level.

## IV. CONCLUSION

A novel wireless sensor design capable of distinctly detecting changes in the ionic content of water and the insertion of solid objects in dielectric container is introduced. The sensor detects changes as low as $3.125 \times 10^{-3}$ M in the ionic content of water and distinguishes between the responses related to the alteration in the concentration of water and insertion of solid objects. Measurements were carried out and results agreed with the expectations from numerical solutions. The application of the sensor in this particular case study can greatly improve the water conservation in public bathrooms and shopping centers. Moreover, its wireless functionality and ability to sense through non-metallic walls position it as a versatile tool for applications in pharmaceutical, chemical, and food industries. The strategic use of array configuration, coupled with beam steering and advanced signal processing methods, heralds a promising future for the widespread adoption of such innovative sensors.

## ACKNOWLEDGMENTS

The authors gratefully acknowledge the support of TUBITAK's Industry Innovation Network Mechanism (SAYEM) Program under grant no. 121D010 (Akıllı Ev Platformu) in collaboration with Eczacıbaşı. H.V.Demir also acknowledges support from TUBA and TUBITAK 2247-A National Leader Researchers Program (121C266). All authors thank to Mr. Fatih Gerenli and Eczacıbaşı team members for fruitfull discussions on smart bathroom applications and further thank Eczacıbaşı for providing us with a sample of ceramic container.

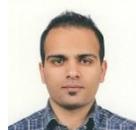
**Sobhan Gholami** joined the Bilkent University Department of Electrical and Electronics Engineering in 2021 as a research graduate student. Prior to his graduate studies, he was an undergraduate researcher at the bioMEMs laboratory of the Department of Electrical and Electronics Engineering at Marmara University,Istanbul-Turkey where he completed his undergraduate studies in 2021. He is currently pursuing his MSc. degree focusing on microwave sensors. His research interests are in electromagnetic theory and application, antenna/microwave theory and measurement.

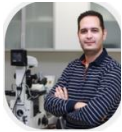 **Emre Unal** received the B.Sc. degree in electrical and electronics engineering from Hacettepe University, Ankara, Turkey, in 2005. He is a full-time Research Engineer at the Institute of Materials Science and Nanotechnology, Bilkent University, Ankara, under the supervision of Prof. H. V. Demir, where he is working on the development of microwave and optoelectronic devices.

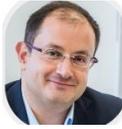 **Hilmi Volkan Demir** (M'04–FM'21) received the B.Sc. degree in electrical and electronics engineering from Bilkent University, Ankara, Turkey, in 1998, and the M.Sc. and Ph.D. degrees in electrical engineering from Stanford University, Stanford, CA, USA, in 2000 and 2004, respectively. In 2004, he joined Bilkent University, where he is currently a professor with joint appointments with the Department of Electrical and Electronics Engineering, Department of Physics and Institute of Materials Science and Nanotechnology (UNAM). He is also a Fellow of the National Research Foundation in Singapore and a Professor with Nanyang Technological University. His research interests include the development of innovative optoelectronic and RF devices. He is a member of Turkish Academy of Science.